%% file: main.tex
\documentclass[aps,prl,twocolumn,amsart,amsmath,amssymb,floatfix,longbibliography]{revtex4-2}
\usepackage{color}
\usepackage{mathrsfs}
\usepackage{graphicx}
\usepackage{dcolumn}
\usepackage{bm}
\usepackage[breaklinks=true, pdftitle={}]{hyperref}
\usepackage[normalem]{ulem}
\usepackage{amsmath,amsfonts,amssymb,ulem}
\usepackage{epstopdf}
\usepackage{xcolor}
\usepackage[separate-uncertainty]{siunitx}

\newcommand{\ma}[1]{{\color{black}#1}}
\newcommand{\re}[1]{{\color{black}#1}}

\usepackage{graphicx}
\begin{document}

\title{Electron slingshot acceleration in relativistic preturbulent shocks explored \\via emitted photon polarization}
%\title{Electron slingshot acceleration in relativistic preturbulent shock explored \\via emitted photon polarization}

\author{Zheng Gong}
%\curraddr{\itshape Current address}:
\email{gong@mpi-hd.mpg.de}\
\thanks{Present address: Department of Mechanical Engineering, Stanford University, Stanford, California 94305, USA. zgong92@stanford.edu}
%\email{\\ asdf}
%\curraddr{sas}
%\curraddr{Present address: Department of Mechanical Engineering, Stanford University, Stanford, California 94305, USA.}

 \affiliation{Max-Planck-Institut f\"{u}r Kernphysik, Saupfercheckweg 1, 69117 Heidelberg, Germany}
 \author{Xiaofei Shen}
 \affiliation{Max-Planck-Institut f\"{u}r Kernphysik, Saupfercheckweg 1, 69117 Heidelberg, Germany}
\author{Karen Z. Hatsagortsyan}
\email[]{k.hatsagortsyan@mpi-hd.mpg.de}
 \affiliation{Max-Planck-Institut f\"{u}r Kernphysik, Saupfercheckweg 1, 69117 Heidelberg, Germany}
\author{Christoph H. Keitel}
 %\email[]{keitel@mpi-hd.mpg.de}
 \affiliation{Max-Planck-Institut f\"{u}r Kernphysik, Saupfercheckweg 1, 69117 Heidelberg, Germany}

\date{\today}
\begin{abstract}
Transient electron dynamics near the interface of counterstreaming plasmas at the onset of a relativistic collisionless shock (RCS) is investigated using particle-in-cell simulations. We identify a slingshot-like injection process induced by the drifting electric field sustained by the flowing focus of backwards-moving electrons, which is distinct from the well-known stochastic acceleration. The flowing focus signifies the plasma kinetic transition from a preturbulent laminar motion to a chaotic turbulence. We find a characteristic correlation between the electron dynamics in the slingshot acceleration and the photon emission features. In particular, the integrated radiation from the RCS exhibits a counterintuitive non-monotonic dependence of the photon polarization degree on the photon energy, which originates from a polarization degradation of relatively high-energy photons emitted by the slingshot-injected electrons. Our results demonstrate the potential of photon polarization as an essential information source in exploring intricate \re{transient dynamics in RCSs} with relevance for earth-based plasma and astrophysical scenarios.

\end{abstract}
\maketitle
Plasma shocks are characterized by rapid steepening of a nonlinear wave, the eventual overtaking by its rear part, and the irreversible energy transfer to the surrounding particles~\cite{sagdeev1966cooperative,courant1999supersonic,landau2013fluid}. 
They are of extensive interest in various scenarios. 
\re{In laboratory, a nonrelativistic shock can generate multi-MeV ions in plasma-based accelerators~\cite{silva2004proton,ji2008generating,fiuza2012laser,haberberger2012collisionless,zhang2017collisionless,yao2023shock}, improve the thermonuclear gain of inertial confinement fusion~\cite{betti2007shock,perkins2009shock,riconda2011kinetic,scott2022shock}, and provide a platform for investigating astrophysical phenomena~\cite{gregori2012generation,kugland2012self,fox2013filamentation,Weibel_NP2015_huntington,li2019collisionless,ruyer2020growth,fiuza2020electron}.
For astrophysics, the shock formed by supernova remnants offer plausible mechanisms towards understanding the origin of TeV cosmic leptons~\cite{koyama1995evidence,diesing2019spectrum,aguilar2019towards_positron,aguilar2019towards_electron} and galactic PeVatrons~\cite{fang2022evidence}, while RCSs are ubiquitous in pulsar wind nebulae~\cite{kirk2009theory}, active galactic nuclei~\cite{romero2017relativistic}, and gamma-ray bursts (GRBs)~\cite{gamma_ray_burst}.
Recent observations suggest that the RCS prompted afterglow radiation indicates a peculiar long GRB from the merger of a compact binary system~\cite{troja2022nearby,rastinejad2022kilonova,mei2022gigaelectronvolt,yang2022long} rather than the core-collapse of massive stars~\cite{woosley2006supernova}, which restimulates the research interest of relevant RCSs~\cite{grovselj2022microphysics}.

The RCS brewed from GRBs is prone to load with $e^{\pm}$ pairs since the $\gamma$-ray photons ahead of the GRB ejecta turn into $e^{\pm}$ via $\gamma$-$\gamma$ reaction~\cite{thompson2000relativistic,meszaros2001pair,beloborodov2002radiation}. Meanwhile, the magnetization level of the unshocked interstellar medium is so low that the unmagnetized initial condition is generally considered~\cite{gruzinov2001gamma}. 
Here, the magnetized filamentation turbulence, self-generated through the filamentation merging and magnetic loop coalescence~\cite{medvedev1999generation,honda2000collective,silva2003interpenetrating,bell2004turbulent,bret2005characterization,califano2006three,bret2008exact,gong2023electron},
is crucial for determining Weibel-mediated shock microstructures~\cite{weibel1959spontaneously,fiuza2012weibel,ruyer2016analytical,marcowith2016microphysics,grassi2017radiation,zhdankin2017kinetic,lemoine2019physics,Magnetic_field_amp_in_shock_2021}, where electrons, undergoing severe swirling and trace crossing, no longer travel in a quasi-layer form. 
The electrons might experience stochastic acceleration, akin to \textit{Fermi} process~\cite{fermi1949origin,fermi1954}, which has been well recognized as sources of energetic electrons in the universe~\cite{bell1978acceleration,spitkovsky2008particle,petrosian2012stochastic,matsumoto2015stochastic,sironi2015relativistic,matsumoto2017electron,comisso2018particle,zhdankin2019electron,comisso2021pitch,lemoine2022first,amano2022nonthermal}.}
\re{Previous studies of unmagnetized RCSs primarily focus on the electron energization and equipartition between electrons and ions through stochastic acceleration~\cite{milosavljevic2006cosmic,Spitkovsky_2008,haugbolle2011three,plotnikov2013particle,naseri2022electron} where
electrons scatter off the self-generated turbulent magnetic structures~\cite{naseri2018growth,peterson2021magnetic,peterson2022magnetic}.
However, it remains largely unexplored how the plasma transits from the nonturbulent flow to kinetic turbulence} and how this transition impacts the acceleration and radiation features in the RCS.

As a versatile information carrier of multi-messenger astrophysics~\cite{bartos2017multimessenger,meszaros2019multi,komatsu2022new}, photon polarization is critical for measuring the magnetic configuration nearby black holes~\cite{akiyama2021first} and crab nebulae~\cite{bucciantini2023simultaneous}. 
%and for analyzing the particle acceleration in the blazar's jet~\cite{liodakis2022polarized}.
Therefore, the question arises whether the polarization feature of spontaneously emitted photons can be employed to reveal the  mechanism responsible for the turbulence transition in a RCS.

In this letter, we investigate the \re{transient electron dynamics} in the transition to turbulence nearby the counterstreaming interface of an \re{unmagnetized pair-loaded RCS precursor, which is potentially associated with the outflow of GRBs.} %This type of RCS potentially associated with GRBs is prone to load with $e^{\pm}$ pairs since the $\gamma$ rays ahead of the GRB ejecta turn into $e^{\pm}$ via $\gamma$-$\gamma$ reaction~\cite{thompson2000relativistic,meszaros2001pair,beloborodov2002radiation}.} 
We employ PIC simulations to examine the photon emission and observe an anomalous non-monotonic dependence (NMD) of the photon polarization degree on the photon energy. We found that the NMD indicates a specific mechanism of electron acceleration, which we term as slingshot injection, caused by a drifting electric field due to the flowing focus of backwards-moving electrons. 
\re{Our slingshot model could be the essential reason accounting for the long-term directed electron heating via electric fields nearby a RCS precursor~\cite{gedalin2008efficient,kumar2015electron,vanthieghem2022origin}.}
Utilizing Hamiltonian analyses, we elucidate that the backwards-flowing focus marks the plasma transition to a turbulent regime in the RCS, which in the electron's transverse phase space is exhibited as the change from the phase-locked to the phase-slipping dynamics. The NMD photon properties stem from a polarization degradation of relatively high-energy photons emitted by the slingshot-injected electrons. The correlation among the NMD of photon polarization, the slingshot injection, and the backwards-flowing focus emphasizes the importance of the transition region to the turbulence in characterizing the acceleration and radiation in the RCS.

\begin{figure}
\includegraphics[width=0.5\textwidth]{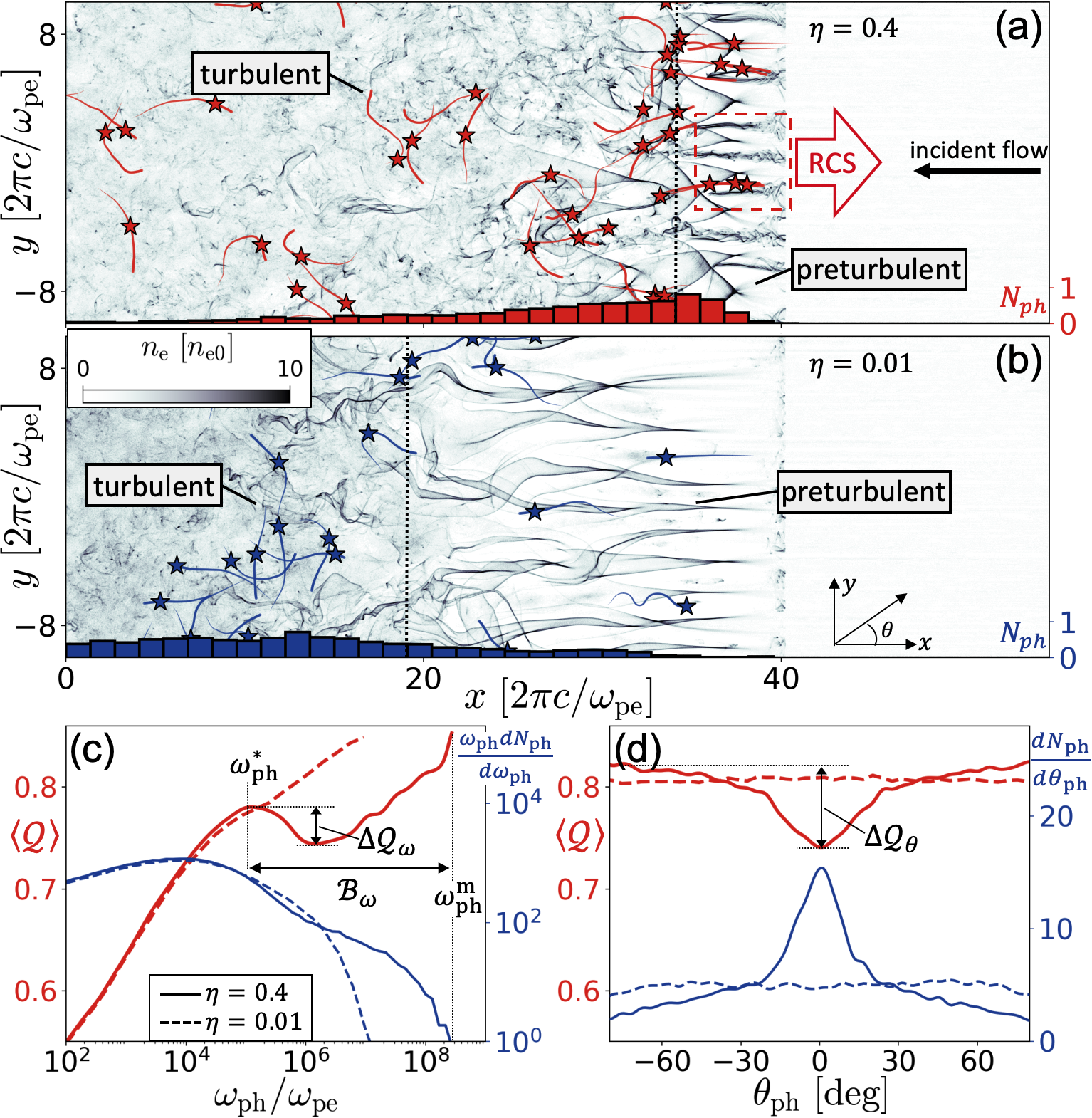}
\caption{The dynamics of a counterstreaming RCS: The electron density $n_e$ at $t=80\pi/\omega_{pe}$ for (a) $\eta=0.4$~\cite{SM_animation_fig_1a} and (b) $\eta=0.01$~\cite{SM_animation_fig_1b}, where lines present the typical electron moving tendency with stars marking the photon emission and the histograms display the spatial distribution of emitted photons with $\omega_{ph}>10^{-2}\omega_{ph}^m$. (c) $\left<\mathcal{Q}\right>$ and $\omega_{ph}dN_{ph}/d\omega_{ph}$ vs $\omega_{ph}$. (d) $\left<\mathcal{Q}\right>$ and $dN_{ph}/d\theta_{ph}$ vs $\theta_{ph}$.}
\label{fig:general_density}
\end{figure}

We have carried out 2D simulations of counterstreaming RCSs, see Fig.~\ref{fig:general_density}. The latter is initiated when a uniform plasma flow with a bulk Lorentz factor $\gamma_0=50$, injected from the right side, is reflected from the left side boundary, which adopts a reflection condition~\cite{spitkovsky2008particle}.
The periodic boundary condition is set in the lateral direction.
%Motivated by the unknown composition of astrophysical jets, 
\re{We consider the flow consisting of electrons, positrons, and ions with the same drifting velocity} and with the number density of $n_{e0}$, $n_{p0}$, and $n_{i0}$, respectively. The charge neutralization $n_{e0}=n_{p0}+Z_i n_{i0}$ is satisfied initially and the ion with charge (mass) $Z_i=1$ ($m_i=1836m_e$) is use. The ratio $\eta \equiv n_{i0}/(n_{i0}+n_{p0})\in(0.01,1)$ denotes the proportion of ions among the whole positive charged particles.
The simulation domain is $200\lambda_{pe}\times 20\lambda_{pe}$ with resolution $\Delta x$=$\Delta y$=$\lambda_{pe}/50$ and $\Delta t=0.95\Delta x/c$. Each cell is filled with 48 macro-particles for each species. 
Here, $\omega_{pe}=(n_{e0}e^2/\varepsilon_0 m_e)^{1/2}$ ($\lambda_{pe}=2\pi c/\omega_{pe}$) is the plasma frequency (skin depth), with the electron charge (mass) $e$ ($m_e$), the vacuum permittivity $\varepsilon_0$,  and the speed of light $c$.
The models of the photon polarization have been implemented in the EPOCH code~\cite{arber2015contemporary,SM}. 
Unless otherwise indicated, we discuss results from the fiducial simulation with $\gamma_0=50$ and $\eta=0.4$.

\begin{figure}[t]
\includegraphics[width=0.97\columnwidth]{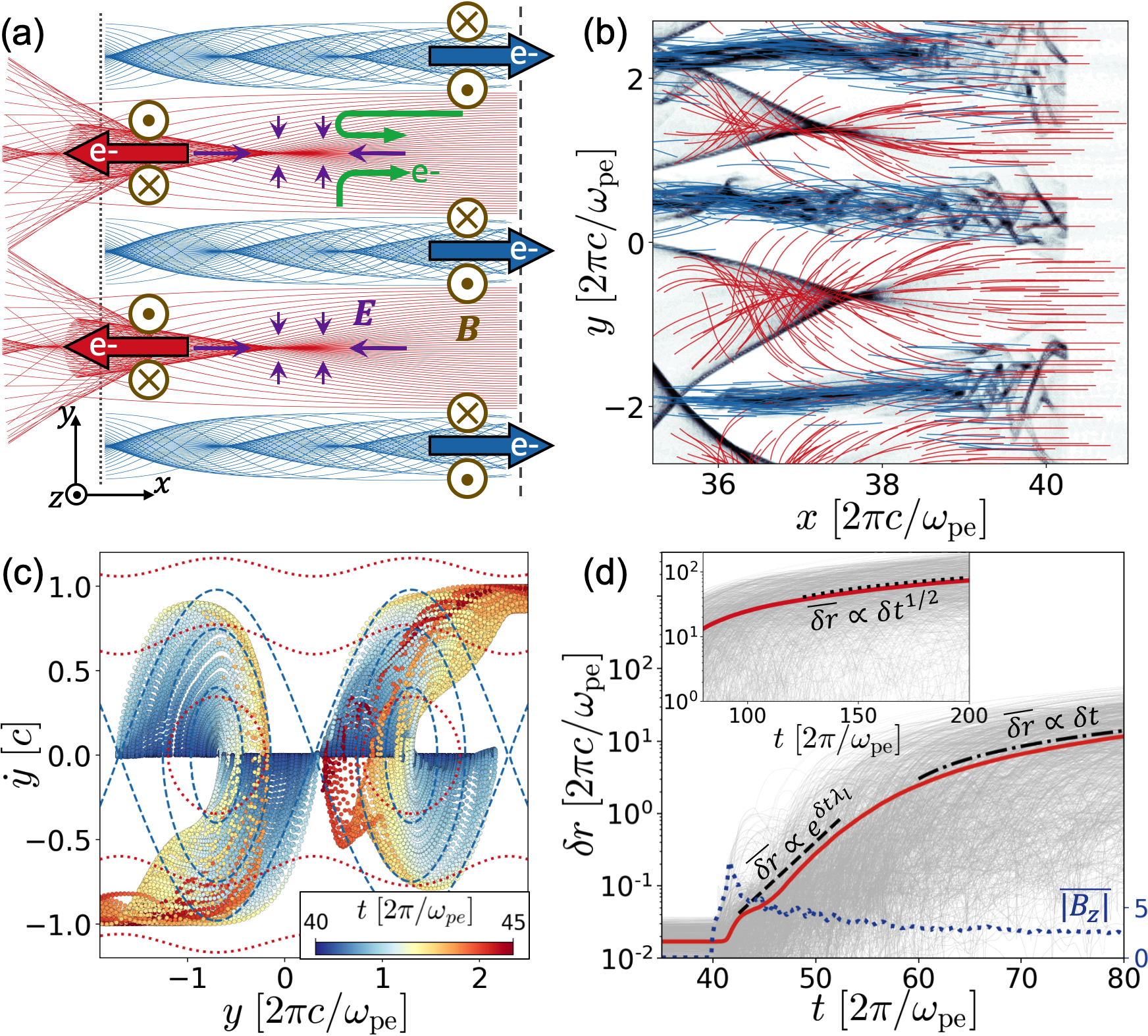}
\caption{(a) Schematic of the backwards-flowing focus with the predicted electron trajectories~\cite{SM_animation_fig_2a}. The brown (purple) markers denote the magnetic (electric) field direction and the green arrows present the slingshot-injected electrons.
(b) Zoom in on the dashed box marked in Fig.~\ref{fig:general_density}(a), where the red (blue) lines represent the backward (forward) moving electrons~\cite{SM_animation_fig_2b}. 
(c) Electron evolution in $(y,\dot{y})$ space with the blue dashed (red dotted) lines contouring $H_\perp|_{\Omega\approx 0.1\omega_{pe}}$ ($H_\perp|_{\Omega^\prime\approx 0.02\omega_{pe}}$)~\cite{SM_animation_fig_2c}.
(d) Time evolution of $\delta r$ ($\overline{\delta r}$) in grey (red). %The simulation results in (b)(c)(d) are calculated by using the same group of electrons.
}
\label{fig:slingshot}
\end{figure}

%The snapshot of the electron density distribution $n_e$ is shown in Fig.~\ref{fig:general_density}(a). It exhibits that the filamentation structure exclusively exists at the front of the RCS interface~\cite{silva2003interpenetrating,bret2008exact}. 
The snapshot of the electron density $n_e$ in Fig.~\ref{fig:general_density}(a) exhibits that the filamentation exclusively exists at the front of the RCS interface. 
Between two adjacent filaments, an electron focusing point emerges, and following that, two oblique density strips stretch out [see Fig.~\ref{fig:slingshot}(b)]. 
Behind the strips, the coherent filaments and focusing points disappear while the turbulence shows up. 
A nontrivial thing is that the photons with energy $\varepsilon_{ph}\equiv \hbar\omega_{ph}> 10^{-2}\hbar\omega_{ph}^m$ are primarily emitted by electrons nearby the interface, where $\omega_{ph}^m\sim 10^8\omega_{pe}$ is the photon cut-off frequency and $\hbar$ the Planck constant.
In contrast, in the case of $\eta=0.01$, the energetic photon emission predominantly occurs in the turbulent region 
%for the case with $\eta=0.01$ 
[see Fig.~\ref{fig:general_density}(b)], even though the preturbulent structures are extended to a larger range.

The degree of photon's linear polarization along the direction of the electron's transverse acceleration is characterized by the Stokes parameter $\mathcal{Q}$~\cite{mcmaster1954polarization}, formulated as~\cite{SM}
\begin{align}\label{eq:Q_theory}
\mathcal{Q}= \frac{\varepsilon_e(\varepsilon_e-\varepsilon_{ph}) K_\frac{2}{3}(\zeta)}{[\varepsilon_e^2+(\varepsilon_e-\varepsilon_{ph})^2]K_\frac{2}{3}(\zeta)-\varepsilon_e(\varepsilon_e-\varepsilon_{ph})\Tilde{K}_\frac{1}{3}(\zeta)},
\end{align}
where $K_{n}(\zeta)$ is the modified secondary Bessel function, $\widetilde{K}_{1/3}(\zeta) = \int_{\zeta}^\infty K_{1/3}(z)\mathrm{d}z$, $\zeta=2\varepsilon_{ph}/[3\chi_e(\varepsilon_e-\varepsilon_{ph})]$, and $\varepsilon_e=\gamma_e m_ec^2$ the electron energy; $\chi_e\equiv(e\hbar/m_e^3c^4)|F_{\mu\nu}p^\nu|$ is the electron quantum strong-field parameter with the field tensor $F_{\mu\nu}$, and the electron four-momentum $p^\nu$. 
%Under the condition of 
At $\chi_e\ll 0.1$, $\partial \mathcal{Q}/\partial \varepsilon_{ph}>0$ predicted by Eq.~\eqref{eq:Q_theory} manifests a monotonic dependence of $\mathcal{Q}$ on $\omega_{ph}$, because for the higher-frequency radiation the formation length is shorter and the preservation of the local polarization degree is improved.
%\ma{understood as that the preservation of the local polarization degree is improved for the higher-frequency radiation with a shorter formation length.}
This monotonic dependence is confirmed by the results of $\eta=0.01$ [see Fig.~\ref{fig:general_density}(c)(d)], where electrons experience stochastic acceleration~\cite{petrosian2012stochastic} and the photon emission is isotropic in the angular space.
However, for $\eta=0.4$ [see Fig.~\ref{fig:general_density}(c)(d)], the averaged polarization degree $\left<\mathcal{Q}\right>$ versus $\omega_{ph}$ exhibits NMD, with a polarization dip $\Delta \mathcal{Q}_\omega\approx 4.5\%$ %compared to the local maximum of $\left<\mathcal{Q}\right>$ 
and a bandwidth ratio $\mathcal{B}_\omega\equiv \omega_{ph}^m/\omega_{ph}^*\sim 10^3$, contradictory to the forementioned monotonic dependence. Here, $\omega_{ph}^*$ is the local maximum point of the function $\left<\mathcal{Q}\right>$ vs $\omega_{ph}$ [see Fig.~\ref{fig:general_density}(c)].
In the angular distribution, $\left<\mathcal{Q}\right>$ has a polarization valley $\Delta\mathcal{Q}_\theta\approx 11\%$ and the photon emission tends to be more collimated within an emission angle $\theta_{ph}\lesssim15^\circ$. 
\re{The features of the slingshot acceleration and the photon NMD are confirmed by 3D PIC simulations~\cite{SM}.}

To unveil the reason of the counterintuitive NMD, we focus on the electron dynamics within the dashed box marked in Fig.~\ref{fig:general_density}(a). 
%For the backwards-moving electron, its transverse deflection is resulted from the transverse inhomogeneity of the forwards-moving charged particles. The simulation results manifest that the quantity of the forward electrons is nearly same as that of the forward ions while the forward positrons are negligible nearby the interface, meaning the effective density approximating $\eta n_{e0}$~\cite{SM}.
For the deflection of backwards-moving electrons nearby the interface, the effective plasma density approximates $\eta n_{e0}$ and the charge density has a sinusoidal profile $\rho\sim |e|\eta n_{e0}\cos[k_y(y-y_c)]$ with $k_y\sim \omega_{pe}/2c$ the periodic wave number and $y_c$ the relative central axis~\cite{SM}.
The self-generated transverse electric and magnetic field is $E_y(y)=(|e| \eta n_{e0}/\varepsilon_0 k_y)\sin[k_y(y-y_c)]$ and $B_z(y)=cE_y$.
As justified by simulations, the energy exchange $d\gamma_e/dt$ is insignificant and thus the transverse dynamics is described by $\ddot{y} + (\Omega^2/k_y)\sin[k_y(y-y_c)] =0$, with $\Omega^2=2\eta n_{e0}e^2/\varepsilon_0\gamma_0 m_e$. 
Then the corresponding Hamiltonian can be derived as~\cite{SM}
\begin{align}\label{eq:slingshot_hami}
H_\perp(y,\dot{y}) = \frac{\Omega^2}{k_y^2}\cos[k_y(y-y_c)] + \frac{1}{2}\dot{y}^2.
\end{align}
Following $H_\perp(y,\dot{y})=H_\perp(y_0,0)$, the electron transverse motion is analyzed as 
\begin{align}\label{eq:slingshot_traj}
t =\frac{k_y}{\sqrt{2}\Omega}\int \frac{dy}{\sqrt{\cos[k_y(y_0-y_c)]-\cos[k_y(y-y_c)] }}.
\end{align}
The trajectories predicted by Eq.~\eqref{eq:slingshot_traj} demonstrate that the backwards-moving electrons would be focused into $y=y_c$ at a restoring time $ t_r\sim 0.6\pi/\Omega$, as confirmed by simulation results [see Figs.~\ref{fig:slingshot}(a)(b)]. 

After the backwards-flowing focus, the electrons start to transit from the preturbulent motion to turbulence, interpreted as a shrinking of the Hamiltonian's separatrix. 
The separatrix $H_\perp(y,\dot{y})\equiv H_\perp(y_c,0)=\Omega^2/k_y^2$ divides the electron dynamics into the confined phase-locked and the escaping phase-slippage regions. 
%The maximum achievable $\dot{y}^*$ is calculated via $H_\perp(y=y_c,\dot{y}=0)=H[k_y(y-y_c)=\pi,\dot{y}^*]$ as $\frac{\dot{y}^*}{c} = \pm 2\frac{\Omega}{ck_y}\sqrt{1-\frac{\Omega^2}{c^2k_y^2}}$.
If the magnetic field decreases with the equivalent restoring frequency reduced from $\Omega$ to $\Omega^\prime$, the phase space volume encompassed by the separatrix is shrunk from $H_\perp(y,\dot{y})<\Omega^2/k_y^2$ to $H_\perp(y,\dot{y})<\Omega^{\prime 2}/k_y^2$. 
Thus, the electrons within the region of $\Omega^{\prime 2}/k_y^2<H_\perp(y,\dot{y})<\Omega^2/k_y^2$ are released into the phase-slippage region [see Fig.~\ref{fig:slingshot}(c)].
%to undergo phase slips [see Fig.~\ref{fig:slingshot}(c)].
The electron release breaks the coherent filament structure and deteriorate the transverse inhomogeneity, leading to the onset of the plasma turbulence.

The transition from the preturbulent flowing focus to the turbulence is illustrated by the evolution of the particle separation [see Fig.~\ref{fig:slingshot}(d)], where $\delta r$ is the distance between an electron and its closest partner at the beginning and $\overline{\delta r}$ refers to the averaged value.
After the focus at $\omega_{pe}t/2\pi \sim 45$, the signature of the chaotic dynamics arises with $\overline{\delta r}\propto \exp{(\lambda_{l} \delta t)}$ characterized by 
%an analogous Lyapunov exponent $\lambda_{l}\approx 0.15\omega_{pe}/\pi$ means that the electrons exhibit a chaotic behavior during the defocus stage~\cite{vulpiani2009chaos,SM_animation_fig_2d_1}
the Lyapunov exponent $\lambda_{l}\approx 0.15\omega_{pe}/\pi$~\cite{vulpiani2009chaos}. The electrons exhibit a chaotic behavior  during the defocusing stage~\cite{SM_animation_fig_2d_1}, where the decrease of the exerted magnetic field $\overline{|B_z|}$ proves the shrinking of the Hamiltonian's separatrix.
Later at $\omega_{pe}t/2\pi\sim 70$, $\overline{\delta r}\propto 0.2c\delta t /\pi $ implies a drifting tendency because of the localized electrons prone to occupy the whole interaction domain~\cite{SM_animation_fig_2d_2}.
%because of the localized electrons prone to occupy the whole interaction domain~\cite{SM_animation_fig_2d_2}. 
Eventually at $ \omega_{pe}t/2\pi >130$, $\overline{\delta r}\propto 9 (\omega_{pe} \delta t/2\pi)^{1/2} $ manifests the electrons' random walk procedure~\cite{ibe2013elements,SM_animation_fig_2d_3}.

\begin{figure}
\includegraphics[width=1\columnwidth]{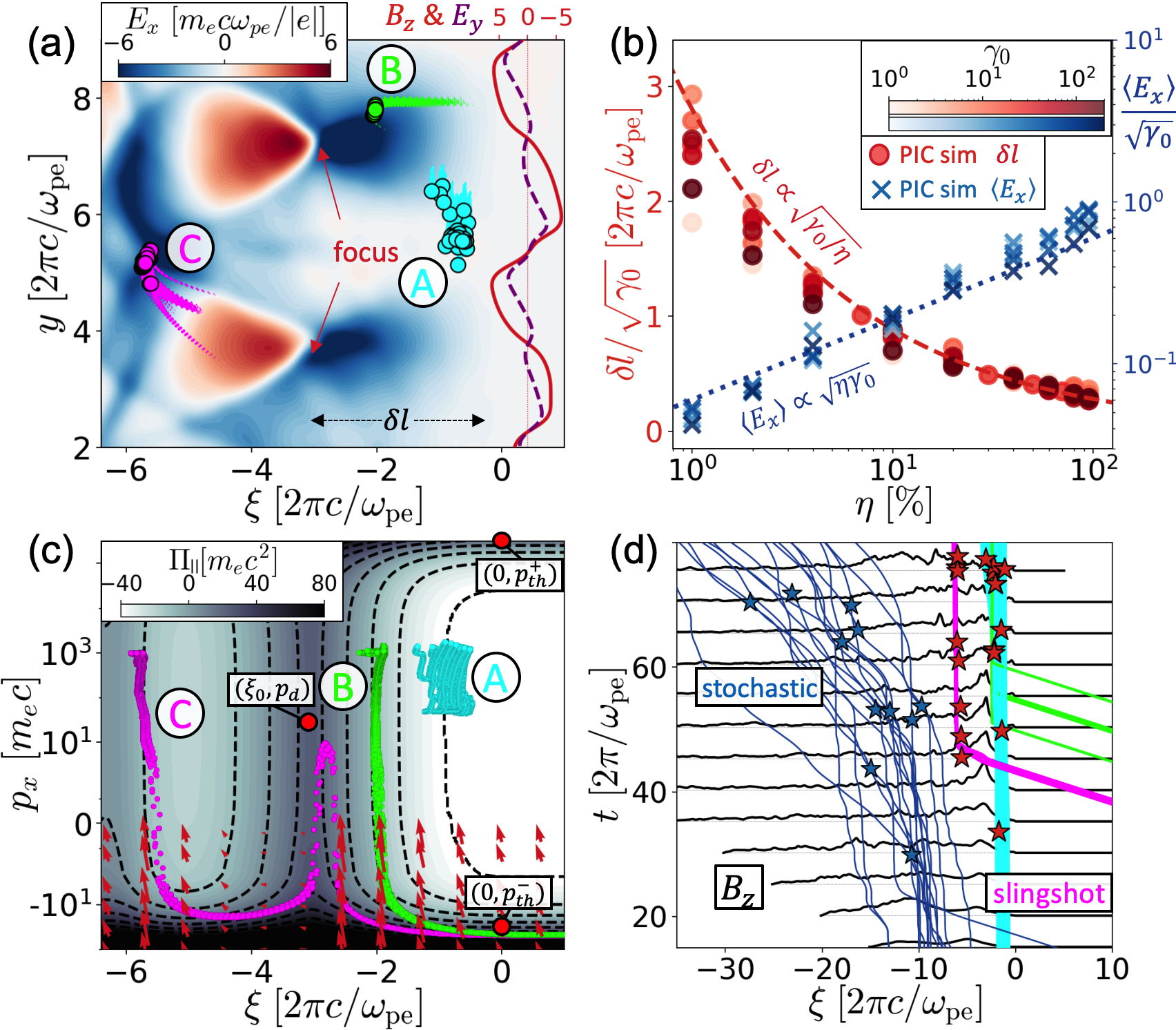}
\caption{(a) Electric field $E_x$ with the transverse profile of $B_z$ and $E_y$.
%(b) $\delta l$ and $\left<E_x\right>$ vs $\eta$ with lines denoting the analytical estimation. 
(b) $\delta l$ and $\left<E_x\right>$ vs $\eta$. 
(c) Hamiltonian $\Pi_\parallel(\xi,p_x)$ with the red arrows denoting the moving tendency modified by the magnetic deflection. 
(d) Time-evolved electron position, where the black lines profile $B_z$ and the red (blue) stars mark the photon emission belong to the slingshot (stochastic) mechanism.
Three kinds of slingshot electrons are shown with the color of `A' cyan, `B' lime, and `C' magenta.}
\label{fig:acc_dynamics}
\end{figure}

The flowing focus leads to a negative longitudinal electric field $E_x$ with a scale length $\delta l$ [see Fig.~\ref{fig:acc_dynamics}(a)], favorable for injecting electrons into the RCS. 
This injection resembles a slingshot, where the filaments serve as the handhold, the backwards-moving electrons behave as the elastic string, and the injected forwards-moving electrons are the projectiles~\cite{SM_animation_shining_eta040}. 
The scale length is calculated as $\delta l\sim t_r c \approx \pi \sqrt{\gamma_0/\eta}(c/\omega_{pe})$.
Given $\nabla\cdot \bm{E}=\rho/\varepsilon_0$, the field strength is estimated as $\left<E_x\right>\approx \pi \sqrt{\eta \gamma_0}m_ec\omega_{pe}/|e|$ [see Fig.~\ref{fig:acc_dynamics}(b)].
The flowing focus successively occurs for the replenished backwards-moving electrons and the field $E_x$ propagates with a velocity $v_x\approx v_d= (1-1/\gamma_0^2)^{1/2}$.
In the interface's co-moving frame $\xi\equiv x-v_d t$, the electron's longitudinal dynamics is determined by the Hamiltonian $\Pi_\parallel(\xi,p_x)=-|e|\varphi(\xi) + c\sqrt{m_e^2c^2+p_x^2}-v_dp_x$ with $\varphi(\xi)=-\int E_x(\xi) d\xi$ [see Fig.~\ref{fig:acc_dynamics}(c)]~\cite{SM}. 
%Considering the separatrix $(\xi_0,p_d)$ with $E_x(\xi_0)=0$ and $p_d=m_ev_d/(1-v_d^2/c^2)^{1/2}$, the relation $\Pi_\parallel(0,p_{th}^\pm)= \Pi_\parallel(\xi_0,p_d)$ governs the injection threshold $p_{th}^-$ and 
Then the injection threshold $p_{th}^-$ and the maximum achievable momentum $p_{th}^+$, derived as~\cite{SM}
\begin{eqnarray}\label{eq:p_max}
p_{th}^{+}\sim 2 \gamma_0^3 + \frac{3}{2}\gamma_0 - \frac{2}{\gamma_0}\ \ \ \&\ \ \ 
p_{th}^{-}\sim -\frac{\gamma_0}{2} - \frac{1}{2\gamma_0}.
\end{eqnarray}
Specifically, there are three types of slingshot-injected electrons [see Figs.~\ref{fig:acc_dynamics}]~\cite{SM_animation_fig_3a}. 
The `A' electrons co-moving with $E_x$ get a pronounced energy gain up to $\gamma_e\sim 10^3$. % for the considered parameters. 
The initially backwards-moving `B' electrons are below the threshold, i.e. $p_x\sim -\gamma_0<p_{th}^-$, but they are still injected because the magnetic deflection $\bm{v}\times\bm{B}$ leading to an attractor effect in $(\xi,p_x)$ space~\cite{hirsch2012differential}, which drags the electrons towards the degraded Hamiltonian $\Pi_\parallel$ [see the red arrows in Fig.~\ref{fig:acc_dynamics}(c)]~\cite{SM}.
The `C' electrons are trapped by the $E_x$ induced by assembling two stretched-out density strips behind the flowing focus position. 
\re{The slingshot acceleration is distinguishable from the previously identified stochastic acceleration, where electrons tend to be repetitively rebounded by magnetic turbulence~\cite{naseri2018growth,peterson2021magnetic,peterson2022magnetic} and undergo Fermi-like stochastic energization~\cite{milosavljevic2006cosmic,Spitkovsky_2008,haugbolle2011three,plotnikov2013particle,naseri2022electron,naseri2018growth}.}
%\re{Unlike the directed slingshot acceleration, the stochastic electrons tend to be repetitively rebounded by the magnetic turbulence and undergo Fermi-like discontinuous acceleration~\cite{spitkovsky2008particle,matsumoto2015stochastic}.}
Figure~\ref{fig:acc_dynamics}(d) manifests that the primary contribution of photon emission nearby the preturbulent interface originates from the slingshot electrons.
\re{The percentage of electrons undergoing a slingshot process is $0.39\%$}.

\begin{figure}
\includegraphics[width=1\columnwidth]{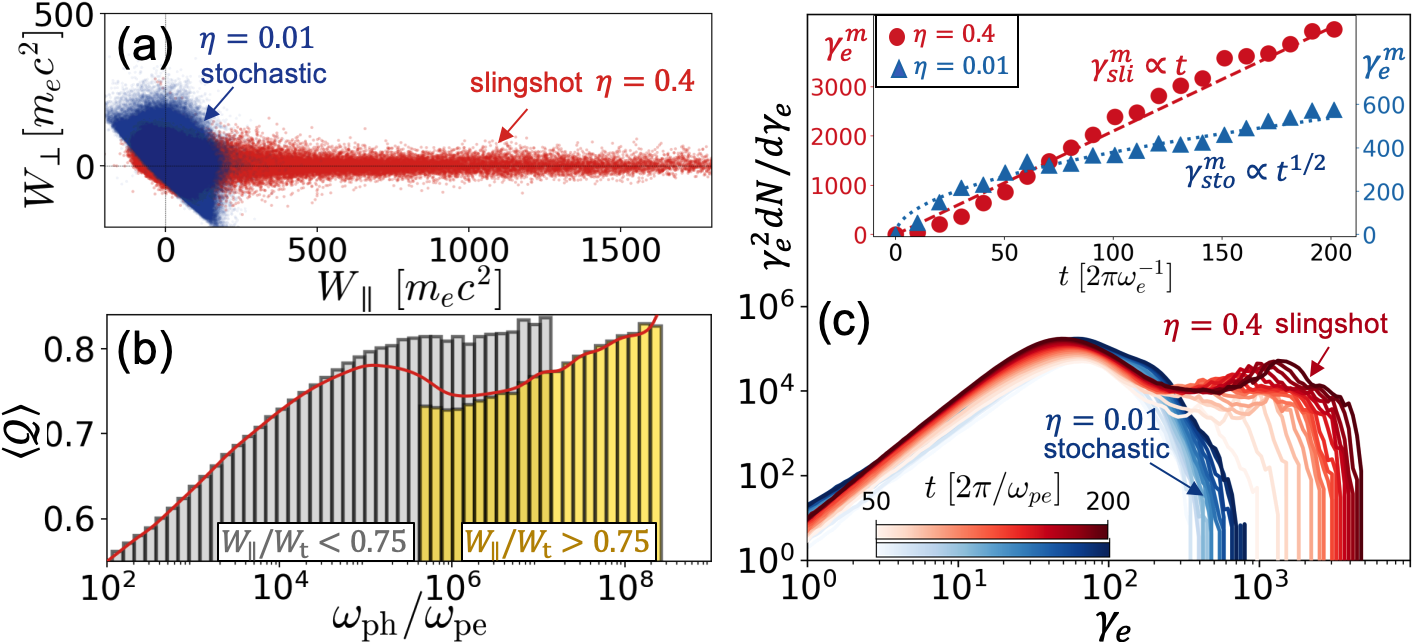}
\caption{(a) Distribution of $W_{\parallel,\perp}$ at the emission moment. (b) $\left<\mathcal{Q}\right>$ vs $\omega_{ph}$ for photons associated with $W_\parallel/W_\mathrm{t}\lessgtr0.75$, where the red line reproduces the $\left<\mathcal{Q}\right>$ vs $\omega_{ph}$ in Fig.~\ref{fig:general_density}(c).
(c) Time-evolved electron energy spectral $\gamma_e^2dN/d\gamma_e$. The inset displays the time-dependent electron maximum energy $\gamma_e^m$. %{\color{red} Maybe show separately polarization $\left<\mathcal{Q}\right>$ of the stochastic accelerated and the slingshot injected electrons? as well as $W_\parallel/W_\mathrm{t}$ for them? Do we need to show the bottom panel $\mathcal{Q}$ in (a)?} 
}
\label{fig:ph_properties}
\end{figure}

%\ma{Another distinct feature between the slingshot and stochastic electrons is the work contribution $W_{\parallel,\perp}$ [see Fig.~\ref{fig:ph_properties}(a)], 
In the search for a criterion distinguishing between the slingshot and stochastic electrons, we turn to the electron's longitudinal and transverse work $W_{\parallel,\perp}$ [see Fig.~\ref{fig:ph_properties}(a)], where $W_\parallel=-\int|e|E_x dx $, $W_\perp=-\int|e|E_y dy $, and $W_\mathrm{t}=W_\parallel + W_\perp$; the integrals are calculated from the beginning to the photon emitting moment.
The slingshot acceleration relies on $E_x$ while the stochastic process is isotropic, meaning that the photon emission associated with $W_\parallel/W_\mathrm{t}\rightarrow 1$ ($W_\parallel/W_\mathrm{t}\rightarrow 0.5$) belong to the slingshot (stochastic) mechanism~\cite{SM_work}.
Therefore, the condition of $W_\parallel/W_\mathrm{t}\lessgtr 0.75$ is a reasonable criterion to distinguish the photon emission from the stochastic or slingshot mechanism. 
For the photons produced from the two mechanisms, both of their $\left<\mathcal{Q}\right>$ vs $\omega_{ph}$ [in Fig.~\ref{fig:ph_properties}(b)] is monotonically increasing as predicted by Eq.\eqref{eq:Q_theory}. 
However, the photon emission from the slingshot is shifted to the higher frequency range compared with the stochastic scenario due to the enhanced energy of slingshot electrons [see Fig.~\ref{fig:ph_properties}(c)]. Therefore, the NMD of $\left<Q\right>$ vs $\omega_{ph}$ comes from the combination between the high polarization degree stochastic photons and the low polarization degree slingshot photons around $\omega_{ph}\sim 10^6\omega_{pe}$ [see Fig.~\ref{fig:ph_properties}(b)]. Nearby this frequency region, the emission of both mechanisms contributes.

\begin{figure}[t]
\includegraphics[width=1\columnwidth]{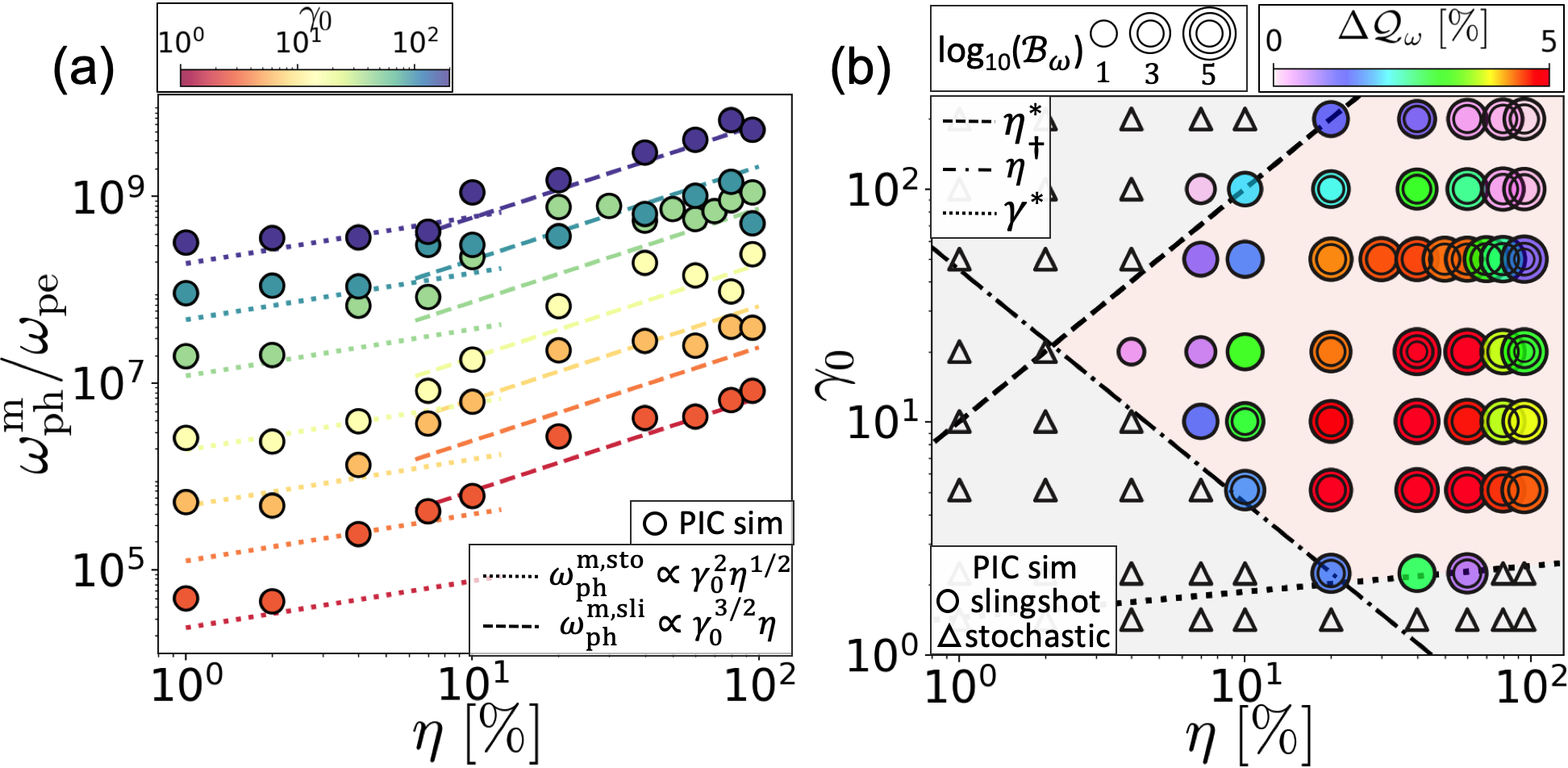}
\caption{(a) $\omega_{ph}^{m}$ vs $\eta$.
(b) Dependence of $\Delta\mathcal{Q}_\omega$ and $\mathcal{B}_\omega$ on $\eta$ and $\gamma_0$, where the circles (triangles) refer to the electron dynamics dominated by the slingshot (stochastic) mechanism. 
%The dependence of (c) $\Delta\mathcal{Q}_\omega$ and (d) $\Delta\omega$ on $\eta$ and $\gamma_0$.
%The lines present the analytically derived dependence or criteria.
}
\label{fig:para_scan}
\end{figure}

%We further estimate the emission characteristics. 
%\re{The slingshot mechanism boosting electrons to $\gamma_e\sim10^3$ much faster than the stochastic one [Fig.~\ref{fig:ph_properties}(c)].  }
%When the electron energy is far from the saturation $\gamma_e\ll p_{th}^+\sim 10^6$, the maximum slingshot energy $\gamma_{sli}^m$ approximates $\gamma_{sli}^{m} \sim |e|\left<E_x\right>\delta t/m_ec\sim \pi \sqrt{\eta \gamma_0}\omega_{pe} \delta t$. 
The maximum slingshot energy $\gamma_{sli}^m$ is approximated as $\gamma_{sli}^{m} \sim |e|\left<E_x\right>\delta t/m_ec\sim \pi \sqrt{\eta \gamma_0}\omega_{pe} \delta t$ when the electron energy is far from the saturation $\gamma_e\ll p_{th}^+\sim 10^6$.
The energy gain of the stochastic process is estimated using the random walk model~\cite{ibe2013elements}: $\gamma_{sto}^{m} \sim 0.5 (\omega_{pe,0} \delta t)^{1/2} \gamma_0^{3/4}\eta^{1/4}$~\cite{SM}. 
These estimates agree well with the simulation results [Fig.~\ref{fig:ph_properties}(c)].
Following $\gamma_{sli}^{m}$ and $\gamma_{sto}^{m}$, the photon cut-off frequency $\omega_{ph}^m$ for the slingshot and stochastic mechanisms is predicted as $\omega_{ph}^{m,sli} \sim \gamma_{sli}^{m\,2}B \propto \gamma_0^{3/2} \eta $ and $\omega_{ph}^{m,sto} \sim \gamma_{sto}^{m\,2}B \propto\gamma_0^{2} \eta^{1/2}$ [see Fig.~\ref{fig:para_scan}(a)] with the magnetic field strength $B\propto \gamma_0^{1/2}$.

Examining the NMD polarization features of the polarization dip $\Delta \mathcal{Q}_\omega$ and the bandwidth $\mathcal{B}_\omega$, we conclude that the high-frequency photon emission is dominated by the slingshot mechanism due to fulfilling three criteria:
i) the photon cut-off frequency originating from slingshot electrons is much higher than via the stochastic mechanism, i.e. $\omega_{ph}^{m,sli}\gg\omega_{ph}^{m,sto}$, reformulated as $\eta\gtrsim \eta^*=0.01\gamma_0$; 
%ii) the number of the slingshot injected electron $N_e^{sli}\sim \left<E_x\right>\delta S$ should be comparable with the most energetic part of the stochastic electrons $N_e^{sto}\sim n_{pe0}L\delta S$, rearranged as $\eta \gtrsim\eta^\dagger \sim0.4\gamma_0^{-1}$, where $\delta S$ is the transverse spatial size and $L$ ;
ii) the number of the slingshot injected electron $N_e^{sli}\propto \left<E_x\right>$ should be larger than the most energetic part of the stochastic electrons $N_e^{sto}\propto n_{pe0}$, rearranged as $\eta \gtrsim\eta^\dagger \propto \gamma_0^{-1}$;
iii) the saturation of the slingshot acceleration should be higher than the stochastic acceleration, i.e. $p_{th}^+\gtrsim \gamma_{sto}^m$, expressed as $\gamma_0\gtrsim\gamma^*= 2\eta^{1/9}$. 
The criteria of the slingshot dominance predicted by $\eta>\mathrm{max}\left\{\eta^*,\eta^\dagger\right\}$ and $\gamma>\gamma^*$ agrees well with the simulation results [see Fig.~\ref{fig:para_scan}(b)]. 
The dependence of $\Delta \mathcal{Q}_\omega$ and $\mathcal{B}_\omega$ on $\eta$ and $\gamma_0$ in Fig.~\ref{fig:para_scan}(b) confirms that the NMD of the polarization degree on photon energy is exclusively from the emission dominated by the slingshot mechanism.

In conclusion, inspecting the origin of unexpected polarization features of the photon radiation in the \re{transient preturbulent RCS precursor}, we have identified the electron slingshot-like acceleration mechanism, distinct from the well-known stochastic acceleration~\cite{milosavljevic2006cosmic,Spitkovsky_2008,haugbolle2011three,plotnikov2013particle,naseri2022electron}. %The slingshot injection is induced by the backwards-flowing focus associated with the transition to turbulence in the RCS's counterstreaming interface.
Our results have implications for both laboratory and astrophysical phenomena.
\re{The identified features of the transition region to turbulence, slingshot injection, and the photon polarization dependence could be actualized in laboratory astrophysics by using the combination of high-energy ion~\cite{muggli2020physics} and $e^{\pm}$ beams~\cite{chen2023perspectives}.
%\re{The slingshot procedure with the achievable maximum energy (see Eq.~\ref{eq:p_max}) could provide an alternative mechanism accounting for TeV cosmic electrons~\cite{aguilar2019towards_electron,SM}.
Moreover, the slingshot electrons, escaping from the preturbulent region to enter the magnetic turbulent plasma~\cite{SM}, potentially behave as the pre-stage injection for the subsequent \textit{Fermi} acceleration in long-term evolved RCSs~\cite{blandford1987particle}.} Finally, the nontrivial photon polarization dynamics implies the necessity of revising the retrieval model for astrophysical magnetic configurations based on radiation features~\cite{zhang2016polarization,akiyama2021first,sironi2021coherent,bucciantini2023simultaneous}.

\begin{acknowledgments}
The original version of code EPOCH adapted here is funded by the UK EPSRC grants EP/G054950/1, EP/G056803/1, EP/G055165/1 and EP/ M022463/1. The authors would like to thank Laurent Gremillet, Anatoly Spitkovsky, and Dmitri Uzdensky for the discussion regarding plasma stream instabilities, the initialization of RCS in PIC simulations, and the undetermined composition of astrophysical jets, respectively. The authors also thank the anonymous referees for their useful comments and suggestions. Z. G. thanks Zhi-Qiu Huang for the gained knowledge about the RCS generated following gamma-ray bursts.
\end{acknowledgments}

%\bibliography{aa}
\input{output.bbl}
\end{document}

%% file: output.bbl

%% file: main.bbl
\begin{thebibliography}{115}
\bibitem [{SM_()}]{SM_work}%
  \BibitemOpen
  \href@noop {} {}\bibinfo {howpublished} {In the PIC simulations, the work $W_{\parallel,\perp}$ could be recorded on the photon particle. When a photon is emitted from its parent electron, it inherits the work value $W_{\parallel,\perp}$ from the electron at that moment. Therefore, each photon would have an attribute of its parental electron's work contribution $W_{\parallel,\perp}$. This is convenient for presenting the correlation between the photon polarization and the work contribution (i.e. the acceleration mechanism).}\BibitemShut {Stop}%
\end{thebibliography}
